# Raman-induced oscillation between an atomic and a molecular quantum gas

C. Ryu[1], X. Du, Emek Yesilada[2], Artem M. Dudarev[3], Shoupu Wan, Qian Niu, and D. J. Heinzen

*Dept. of Physics, The University of Texas, 1 University Station C1600, Austin, Texas, 78712, USA*

1. Present address: Physics Division, NIST, Gaithersburg, Maryland 20899-8424, USA

2. Present address: Freescale Semiconductor, 3501 Ed Bluestein Blvd., Austin, TX, 78721, USA

3. Present address: Max-Planck-Institut für Physik komplexer Systeme, Nöthnitzer Str. 38, 01187 Dresden, Germany

**It has recently been demonstrated that quantum degenerate gases of very weakly bound molecules can be produced by atomic gases with Feshbach resonances[1-5]. More strongly bound molecules can be produced with Raman photoassociation of a quantum gas[6-8], although this process has not yet been shown to produce a quantum degenerate molecular gas. In principle, Feshbach resonance and Raman photoassociation can be quantum-mechanically reversible, and lead to collective coherent phenomena such as Rabi cycling between an atomic and a molecular gas[9-11]. However, such atom-molecular coherence has only partly been realized experimentally[12]. Effects that may limit coherence include thermalizing elastic collisions, inelastic collisions, spontaneous Raman scattering[6-8,11], and pairing field formation[13-15]. Here, we demonstrate a method that circumvents these limitations, based on Raman photoassociation of atoms in an optical lattice and driven into a Mott insulator state[16--18], as proposed by Jaksch *et al.*[19]. We find that the Raman photoassociation transition is resolved into discrete lines corresponding to the**

**quantized lattice site occupancies, and demonstrate that this provides a new method to accurately determine the distribution of site occupancies and the atom-molecule scattering length. Furthermore, we observe a Raman-induced oscillation of the central core of the gas, containing about 30% of the atoms, between an atomic and a molecular gas. This is the first demonstration of the quantum-mechanically reversible production of a molecular quantum gas, and of the formation of a molecular quantum gas in more deeply bound states by Raman photoassociation. This result may lead to further studies of molecular quantum gases, to the production of quantum gases of ro-vibrational ground state[19] or heteronuclear[20] molecules, or to new techniques for quantum information processing with atoms in optical lattices[21].**

The recent observation[18] of a quantum phase transition between a superfluid and a Mott insulator state of a Bose gas[16,17] was an important building block of our experiment. Such a transition can be generated when an optical lattice potential of the form $V(x,y,z,t) = V_0(t)[\sin^2(kx) + \sin^2(ky) + \sin^2(kz)]$ is superimposed onto a Bose-condensed gas. This potential arises from the optical dipole force from six interfering laser beams of wavelength $\lambda$ and wavevector $k = 2\pi/\lambda$. As first emphasized by Jaksch *et al.*[17], this system provides an almost perfect laboratory realization of the Bose-Hubbard model[16]. In this model, identical bosons in a lattice evolve according to the Hamiltonian

$$H = \sum_i n_i \varepsilon_i - J \sum a_i^+ a_j + U \sum_i \frac{1}{2} n_i (n_i - 1) \qquad (1)$$

In this expression, the index i labels the lattice sites, $n_i = a_i^+ a_i$ is the atomic number operator for site i, with $a_i^+$ and $a_i$ the raising and lowering operators for site i, and $\varepsilon_i$ is the single particle energy for site i. The second summation is over all distinct pairs of nearest neighbor sites. The term proportional to J accounts for hopping of particles between nearest neighbor sites, which occurs in the experiments by quantum-mechanical tunneling. The term proportional to U gives the contact energy between particles, which arises in the experiments from elastic collisions. At zero temperature, the Bose-Hubbard model exhibits a quantum phase transition from a superfluid state at low values of U/J to a Mott insulator state for large values of U/J. In a homogeneous Mott insulator state, the number of particles in each site is nearly fixed to some identical integer value.

Each cycle of our experiment began with the production of a Bose-Einstein condensate[22,23] of $^{87}$Rb atoms in their $|\,f=1, m_f = -1\rangle$ state, where f and $m_f$ give the total spin and laboratory spin projection quantum numbers of the atom. We could adjust the total number of atoms N to be any value between $1 \times 10^5$ and $2 \times 10^6$, with an initial condensate fraction of at least 80%. The condensate was trapped in a "cloverleaf" type magnetic trap[24] with trapping frequencies of $\omega_r/2\pi = 20.7$ Hz and $\omega_z/2\pi = 11.6$ Hz. We then imposed an optical lattice formed by six TEM$_{00}$-mode Gaussian laser beams of wavelength $\lambda = 830$ nm, arranged in three counter-propagating beam pairs. One beam pair was aligned along the weak (z) axis of the magnetic trap, and had a waist size of 130 μm; the other two beam pairs were aligned perpendicular to the weak axis of the magnetic trap and to each other, and had waist sizes of 260 μm. Interference between the three beam pairs was suppressed by orienting their polarization vectors to be mutually perpendicular and by imposing frequency shifts between them. In the experiments, we ramped the lattice height $V_0$ from 0 to some maximum value $V_{0m}$ in 50 ms, with $V_{0m}$ larger than the value 13.0 $E_R$ at which the superfluid-to-insulator transition occurs for our conditions. Here, $E_R = (\hbar k)^2/2M$ is the single-photon recoil energy, and M is the mass of a Rb atom. We verified that we had reached the Mott insulator phase by reproducing the time-of-flight images first observed by Greiner *et al.*[18].

We then studied the stimulated Raman photoassociation spectrum of the gas prepared by ramping the lattice strength to $V_{0m} = 16.7 \pm 1.1$ $E_R$. After the 50 ms ramp, and with the lattice beams and the magnetic trapping fields left on, we illuminated the gas with two additional coherent laser beams of frequency $\omega_1$ and $\omega_2$. Their respective wavevectors $\mathbf{k}_1$ and $\mathbf{k}_2$ were parallel. As shown in Fig. 1, a colliding pair of atoms of relative energy $\varepsilon_f$ may simultaneously absorb a photon of wavevector $\mathbf{k}_1$, emit a photon of wavevector $\mathbf{k}_2$, and make a transition from its initially free state to a bound state of energy $\varepsilon_b$ relative to the dissociation energy. We tuned the frequency $\omega_1$ by an amount $\Delta/2\pi = 3.9$ GHz to the red of a single-color photoassociation resonance to a $0_g^-$ (J=0) vibrational level at an energy approximately 28 cm$^{-1}$ below the $5^2S_{1/2} + 5^2P_{1/2}$ dissociation limit. We tuned the Raman frequency difference $\omega_R = \omega_2 - \omega_1$ near resonance with the stimulated Raman transition to the second-to-last $\ell = 0$ vibrational level of the $|F=2, M_F = -2\rangle$ adiabatic potential curve which asymptotically connects to the $|f=1, m_f = -1\rangle + |f=1, m_f = -1\rangle$ dissociation limit. This transition has previously been studied in the irreversible case for a condensate[6,8], and a Mott insulator[7]. Here F and $M_F$ give the total spin and laboratory spin projection quantum numbers of the molecule, and $\ell$ is the relative orbital angular momentum of the two atoms in the molecule.

In this experiment, each atom finds itself trapped in the quantum ground state of a single lattice site for the duration of the Raman photoassociation pulse. The lattice potential near the minimum is approximately a spherical harmonic oscillator potential with vibration frequency $\omega_v/2\pi \approx (4E_R V_{0m})^{1/2}/h = 27$ kHz. If an atom singly occupies a lattice site, it cannot undergo photoassociation. If two atoms occupy a lattice site, then their motion is separable into motions of the center-of-mass and of the relative separation $\mathbf{R}$. The relative motion occurs with reduced mass $\mu$ in a spherically symmetric potential $V_{rel}(R)$ consisting of the harmonic trapping potential $(1/2)\mu\omega_v^2 R^2$ added onto the ordinary interatomic potential, as shown in Fig. 1. This has the effect of discretizing the first few energy states above the dissociation limit with a level separation $\hbar\omega_v$.

We first measured the fraction of atoms $N(\tau)/N(0)$ remaining in the gas as a function of the Raman difference frequency $\omega_R$, for a fixed Raman pulse duration $\tau = 30$ ms and for Raman laser intensities $I_1 = 11.7$ W/cm$^2$ and $I_2 = 6.5$ W/cm$^2$. The number of atoms was determined by measuring the total absorption by the atoms of a probe laser pulse tuned near an atomic transition resonance. Typical results are shown in Fig. 2a for $N = 4.7 \times 10^5$ atoms, and in Fig. 2b for $N = 2.4 \times 10^5$ atoms. Both spectra show a resonant depletion of atoms from the gas when the Raman frequency $\omega_R$ is tuned near $\varepsilon_b/\hbar$. The spectrum of Fig. 2a is split into two peaks, whereas that of Fig. 2b has a single peak.

Because our gas is confined in a weak harmonic trap, its density is highest in the center and falls to zero at the edges. With the lattice on, with $V_{0m} = 16.7$ $E_R$, and with a central site occupancy $n_c \geq 1$ atom per site, the gas tends locally to be in an insulating state of the Bose-Hubbard model, and therefore to have a nearly quantized site occupancy, but that occupancy varies with position[17]. If $n_c = 2$, a core of the gas with 2 atoms per site is surrounded by a shell with 1 atom per site. Taking into account the Gaussian profiles of the lattice laser beam dipole forces, we estimate that this occurs in our experiment for total atom number between about $1.5 \times 10^5$ and $5 \times 10^5$. For higher atom number, we expect that $n_c = 3$, so that a core of the gas with 3 atoms per site is surrounded by a shell with 2 atoms per site, which in turn is surrounded by another shell with 1 atom per site.

We conclude that the resonance peak in Fig. 2b and the lower frequency peak in Fig. 2a arise from the photoassociation of two atoms in one lattice site to produce one molecule, and that the higher frequency peak in Fig. 2a arises from the photoassociation of three atoms in one site to make one atom and one molecule. As shown in Fig. 3, the transition for these two cases is not expected to occur at a Raman frequency of $\omega_R = \varepsilon_b/\hbar$ but rather at frequencies of

$$\omega_{R2} = \frac{\varepsilon_b'}{\hbar} + \frac{1}{\hbar}\left[2E_{La} - E_{Lm} + 3\hbar\omega_{va} + U_{aa} - \frac{3}{2}\hbar\omega_{vm}\right] \approx \frac{\varepsilon_b'}{\hbar} + \frac{1}{\hbar}\left[\frac{3}{2}\hbar\omega_v + U_{aa}\right] \quad (2a)$$

$$\omega_{R3} = \frac{\varepsilon_b'}{\hbar} + \frac{1}{\hbar}\left[2E_{La} - E_{Lm} + \left(\frac{9}{2}\hbar\omega_{va} + 3U_{aa}\right) - \left(\frac{3}{2}\hbar\omega_{va} + \frac{3}{2}\hbar\omega_{vm} + U_{am}\right)\right]$$
$$\approx \frac{\varepsilon_b'}{\hbar} + \frac{1}{\hbar}\left[\frac{3}{2}\hbar\omega_v + 3U_{aa} - U_{am}\right] \quad (2b)$$

for the 2 and 3 atom resonances, respectively. Here, $\varepsilon_b' = \varepsilon_b + E_{AC}$, where $E_{AC}$ is the AC stark shift of the Raman transition by the photoassociation lasers, not shown in Fig. 3. $E_{La}$ and $E_{Lm}$ are the AC Stark shifts of the atom and molecule from the lattice laser beams at the lattice potential minima, and $\omega_{va}$ and $\omega_{vm}$ the vibration frequencies of an atom and molecule in a lattice site, respectively. We have studied the AC stark shifts induced by both the lattice and the photoassociation laser beams, and determined that $E_{Lm} = 2E_{La}$ to within the accuracy of the data of Fig. 2. This implies that $\omega_{vm} = \omega_{va} \equiv \omega_v$, and that the approximate equalities in Eqs. 2 apply. Also, we found that $E_{AC}/h \approx 75$ kHz. The remaining 42 kHz of the shift of the two atom resonance from its unperturbed value of 636.010 kHz [6] is consistent with the value $(3/2)\omega_v/2\pi + U_{aa}/h$.

The splitting between the two peaks of Fig. 2a is $\delta\omega_R = (\omega_{R3} - \omega_{R2}) = (2U_{aa} - U_{am})/\hbar$. $U_{aa}$ is the contact energy between two atoms in one lattice site, and $U_{am}$ is the contact energy between one atom and one molecule in the one lattice site, with

$$U_{aa} = \frac{2\pi\hbar^2 a_{aa}}{M/2}\int d^3x |\psi_a(\vec{x})|^4 = \delta_{aa}\frac{a_{aa}}{a}E_R \quad (3a)$$

$$U_{am} = \frac{2\pi\hbar^2 a_{am}}{2M/3}\int d^3x |\psi_a(\vec{x})|^2|\psi_m(\vec{x})|^2 = \delta_{am}\frac{a_{am}}{a}E_R \quad (3b)$$

where $a = \lambda/2$ is the lattice period, $a_{aa}$ and $a_{am}$ are the atom-atom and atom-molecule scattering lengths, and $\psi_a$ and $\psi_m$ are the single-particle Wannier functions for the atom and molecule, respectively. The atomic scattering length is $a_{aa} = 100.4 \pm 0.1$ $a_0$ [25], where $a_0$ is the Bohr radius. At a lattice depth of $V_{0m} = 16.7\ E_R$, numerical calculations of the Wannier functions give the result $\delta_{aa} = 33.4$ and $\delta_{am} = 41.0$, accounting also for a 2% reduction in $\delta_{aa}$ due to interaction-induced couplings to higher band states of the potential. Together with equations 2-3 above, the measured splitting $\delta\omega_R/2\pi = 3.00 \pm 0.13$ kHz allows us to determine that the atom-molecule scattering length is $a_{am} = -8 \pm 14\ a_0$. The error in this result is much smaller than that of a previous result for the same quantity, $-180 \pm 150\ a_0$ [6], because the lattice discretizes the contact energy shifts and makes the calibration of atomic and molecular density much less uncertain.

The 1.4 kHz width of the 2-atom peak is due primarily to power broadening. The power-broadened width of the 3-atom peak should be about $\sqrt{3}$ times larger, i.e. about 2.4 kHz. Possibly, the observed 3.0 kHz width of this peak contains a contribution from the loss rate of particles in the final state due to inelastic atom-molecule collisions. Further quantitative studies of the linewidths would be required to establish whether this is the case.

On resonance, the 30 ms duration Raman photoassociation pulse depletes more than 95% of the atoms from multiply-occupied sites, assuming that all three atoms are lost from triply occupied sites due to inelastic atom-molecule collisions. Therefore, the Raman photoassociation spectrum provides a quantitative measure of the distribution of site occupancies of the gas. From Fig. 2a, we can deduce that about 49% of the atoms resided in singly occupied sites, 28% in doubly occupied sites, and 23% in triply occupied sites. For the data of Fig. 2b, we deduce that about 33% of the atoms resided in doubly occupied sites and 67% in singly occupied sites. The triply occupied sites appear at a slightly lower atom number than our estimate, but otherwise the distributions are in qualitative agreement with what we expect based on our gas parameters.

We next measured the number of atoms remaining in the gas as a function of the Raman laser pulse duration, with the Raman difference frequency fixed onto resonance with the lower frequency peak, and with $N = 2.8 \times 10^5$, $V_{0m} = 15.0 \pm 1$ $E_R$, $I_1 = 20.0$ W/cm$^2$ and $I_2 = 5.6$ W/cm$^2$. As shown in Fig. 4, the system exhibits a clear, coherent Rabi oscillation. The figure shows a fit to the theoretical state population in a model in which a fraction $F_2$ of the atoms undergoes a coherent oscillation on the atom-molecular transition with a Rabi frequency $\Omega$ and with a molecular population decay rate $\Gamma$. From the fit we obtain $F_2 = 0.308$, $\Omega = 4126$ s$^{-1}$, and $\Gamma = 179$ s$^{-1}$. The finite loss rate $\Gamma$ presumably arises from the inelastic scattering of Raman laser photons by the molecules[6,11].

The implication of this result is that the central core of the gas, containing about 30% of the atoms, oscillates between a Mott insulator state with two atoms per site, and a molecular gas with one molecule per site. At integer plus one-half multiples of the Rabi period $2\pi/\Omega$, the central core of the sample constitutes a quantum degenerate molecular gas. We cannot say whether this gas is a Mott insulator because we do not know the molecule-molecule interaction. At other times, the central core of the gas consists of a coherent superposition of an atom pair and a molecule at each site. The data of Fig. 4 illustrates the highest possible level of "quantum control" of pairs of atoms. Molecules are deterministically produced in a regular array, with every degree of freedom including their spatial coordinate and their ro-vibrational state exactly specified within the limits set by quantum mechanics. This high degree of control results from several factors[18]. The initial Mott insulator state has almost no entropy. Inelastic collisions are suppressed by the isolation of each atom pair in one lattice site. The quantization of the mean field shift eliminates a source of inhomogeneous broadening. Finally, the quantization of the continuum states suppresses generation of excited atoms due to pairing field effects[13-15].

In summary, we have studied the Raman photoassociation of a gas of $^{87}$Rb atoms in a Mott insulator state. Due to atomic interactions, the stimulated Raman spectrum was resolved into separate lines corresponding to the different integer occupancies of the lattice sites. This allowed us to measure the distribution of occupancies of the lattice sites, and to determine the atom-molecule elastic scattering length. With the laser tunings fixed on Raman resonance for doubly occupied sites, we observed a reversible oscillation of the central core of the gas, containing about 30% of the atoms, between an atomic and a molecular quantum gas. This is the first observation of such a high degree of reversibility and coherence in a coupled atom-molecular gas.

1. Herbig, J. *et al.* Preparation of a pure molecular quantum gas. *Science* **301**, 1510 (2003).

2. Xu, K. *et al.* Formation of quantum-degenerate sodium molecules. *Phys. Rev. Lett.* **91**, 210402 (2003).

3. Greiner, M., Regal, C.A., & Jin, D.S. Emergence of a molecular Bose-Einstein condensate from a Fermi Gas. *Nature* **426**, 537 (2003).

4. Zwierlein, M. W. *et al.* Observation of Bose-Einstein condensation of molecules. *Phys. Rev. Lett.* **91**, 250401 (2003).

5. Jochim, S. *et al.* Bose-Einstein condensation of molecules. *Science* **302**, 2101 (2003).

6. Wynar, R., Freeland, R. S., Han, D. J., Ryu, C., & Heinzen, D. J. Molecules in a Bose-Einstein Condensate. *Science* **287**, 1016 (2000).

7. Rom, T. *et al.* State-Selective Production of Molecules in Optical Lattices. *Phys. Rev. Lett.* **93**, 073002 (2004).

8. Thalhammer, G., Theis, M., Winkler, K., Grimm, & R., Denschlag, J. H. Inducing an optical Feshbach resonance via stimulated Raman scattering. *Phys. Rev. A* **71**, 033403 (2005).

9. Timmermans, E., Tommasini, P., Hussein, M., & and Kerman, A. Feshbach resonances in atomic Bose-Einstein condensates. *Phys. Rep.* **315**, 199-230 (1999).

10. Javanainen, J., & Mackie, M. Coherent photoassociation of a Bose-Einstein condensate. *Phys. Rev. A* **59**, R3186-3189 (1999).

11. Heinzen, D. J., Wynar, R. H., Drummond, P. D., & Kheruntsyan, K. V. Super-Chemistry: Coherent Dynamics of Coupled Atomic and Molecular Bose Condensates. *Phys. Rev. Lett.* **84**, 5029 (2000).

12. Donley, E. A., Claussen, N. R., Thompson, S. T., & Wieman, C. E. Atom-molecule coherence in a Bose-Einstein condensate. *Nature* **417**, 529 (2002).

13. Holland, M., Park, J., & Walser, R. Formation of Pairing Fields in Resonantly Coupled Atomic and Molecular Bose-Einstein Condensates. *Phys. Rev. Lett.* **86**, 1915 (2001).

14. Kokkelmans, S. J. J. M. F, & Holland, M. Ramsey fringes in a Bose-Einstein condensate between atoms and molecules. *Phys. Rev. Lett.* **89**, 180401 (2002).

15. Köhler, T., Gasenzer, T., & Burnett, K. Microscopic theory of atom-molecule oscillations in a Bose-Einstein condensate. *Phys. Rev. A* **67**, 013601 (2003).

16. Fisher, M. P. A., Wiechman, P. B., Grinstein, G., & Fisher, D. S. Boson localization and the superfluid-insulator transition. *Phys. Rev. B* **40**, 546 (1989).

17. Jaksch, D., Bruder, C., Cirac, J. I., Gardiner, C. W., & Zoller, P. Cold Bosonic Atoms in Optical Lattices. *Phys. Rev. Lett*. **81**, 3108-3111 (1998).

18. Greiner, M., Mandel, O., Esslinger, T., Hänsch, T. W., & Bloch, I. Quantum phase transition from a superfluid to a Mott insulator in a gas of ultracold atoms. *Nature* **415**, 39 (2002).

19. Jaksch, D., Venturi, V., Cirac, J. I., Williams, C. J., & Zoller, P. Creation of a molecular condensate by dynamically melting a Mott Insulator. *Phys. Rev. Lett.* **89**, 040402 (2002).

20. Damski, B. *et al.* Creation of a Dipolar Superfluid in Optical Lattices. *Phys. Rev. Lett.* **90**, 110401 (2003).

21. Jané, E., Vidal, G., Dür, W., Zoller, P., & and Cirac, J. I. Simulation of quantum dynamics with quantum optical systems. *Quant. Inf. Comp. 3*, 15 (2003).

22. Anderson, M. H., Ensher, J. R., Matthews, M. R., Wieman, C. E., & Cornell, E. A. Observation of Bose-Einstein Condensation in a Dilute Atomic Vapor. *Science* **269,** 198-201 (1995).

23. Davis, K. B. *et al.* Bose-Einstein Condensation in a Gas of Sodium Atoms. *Phys. Rev. Lett.* **75**, 3969-3973 (1995).

24. Mewes, M.-O. *et al.* Bose-Einstein Condensation in a Tightly Confining dc Magnetic Trap. *Phys. Rev. Lett.* **77**, 416-419 (1996).

25. van Kempen, E. G. M. *et al.* Inter-Isotope Determination of Ultracold Rb Interactions from Three High-Precision Experiments. *Phys. Rev. Lett.* **88**, 093201 (2002).

**Acknowledgements**

We gratefully acknowledge the support of this work by the National Science Foundation, the R.A. Welch Foundation, the Fondren Foundation, and the ONR Quantum Optics Initiative.

**Competing interests statement**

The authors declare that they have no competing financial interests.

Correspondence and requests for materials should be addressed to DH (heinzen@physics.utexas.edu)

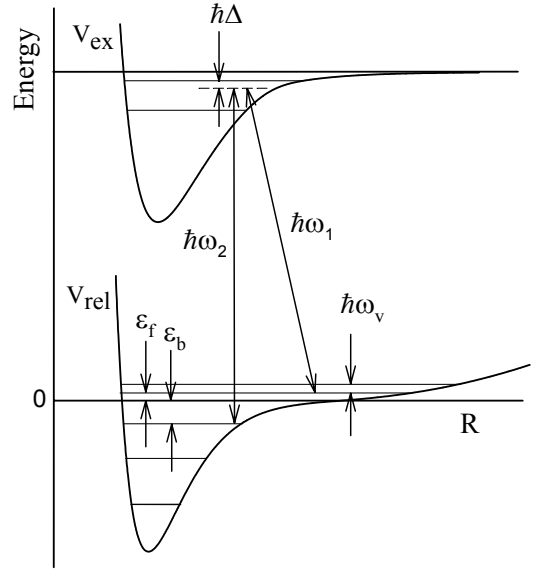

**Figure 1** Raman photoassociation of two atoms in an optical lattice site.

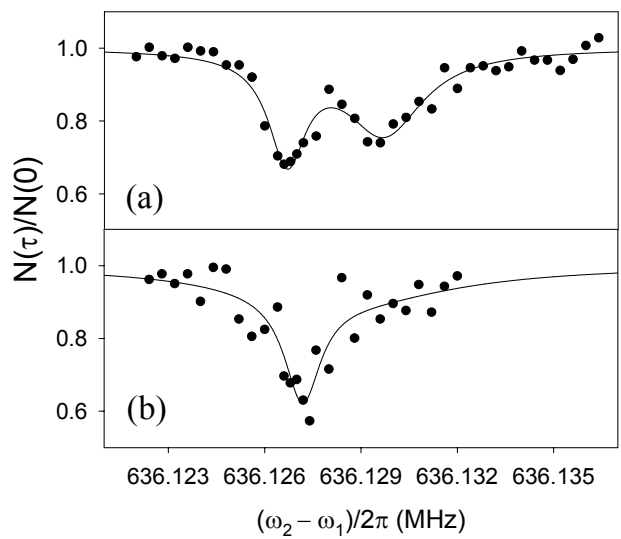

**Figure 2** Raman photoassociation spectrum of atoms in an optical lattice for a total atom number (a) $N = 4.7 \times 10^5$, and (b) $N = 2.4 \times 10^5$.

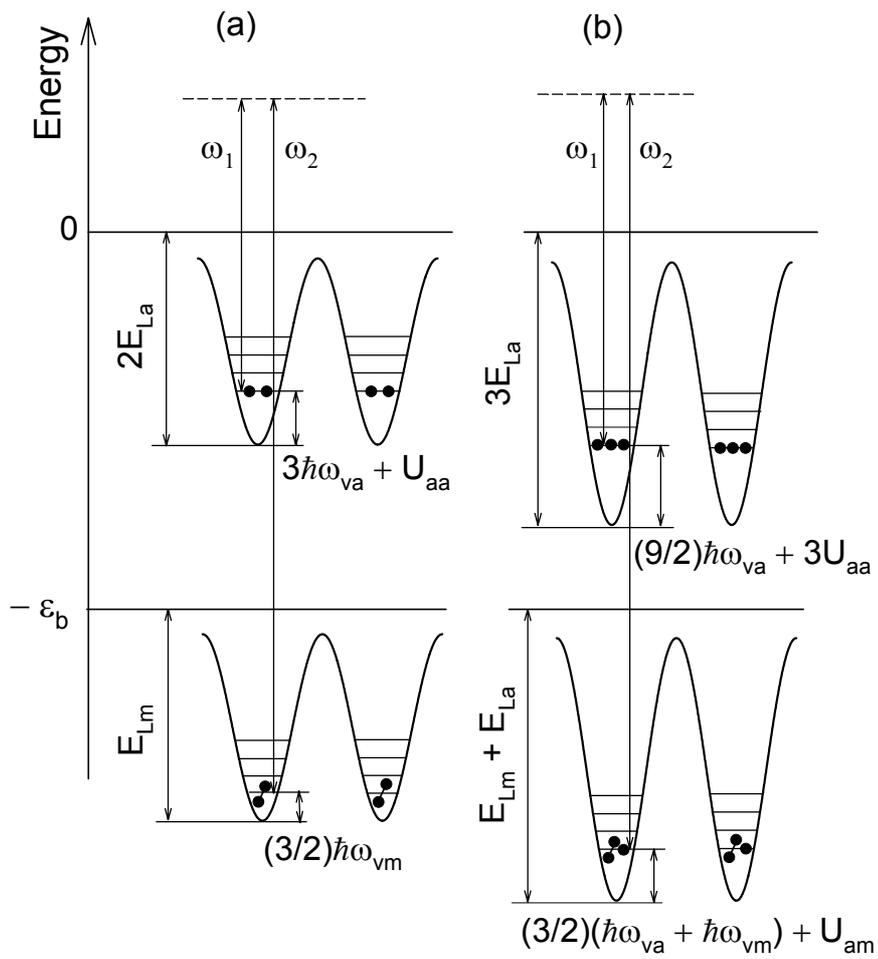

**Figure 3** Energy diagram illustrating (a) the Raman photoassociation of two atoms in one lattice site to make one molecule, and (b) the Raman photoassociation of three atoms in one lattice site to make one molecule and one atom.

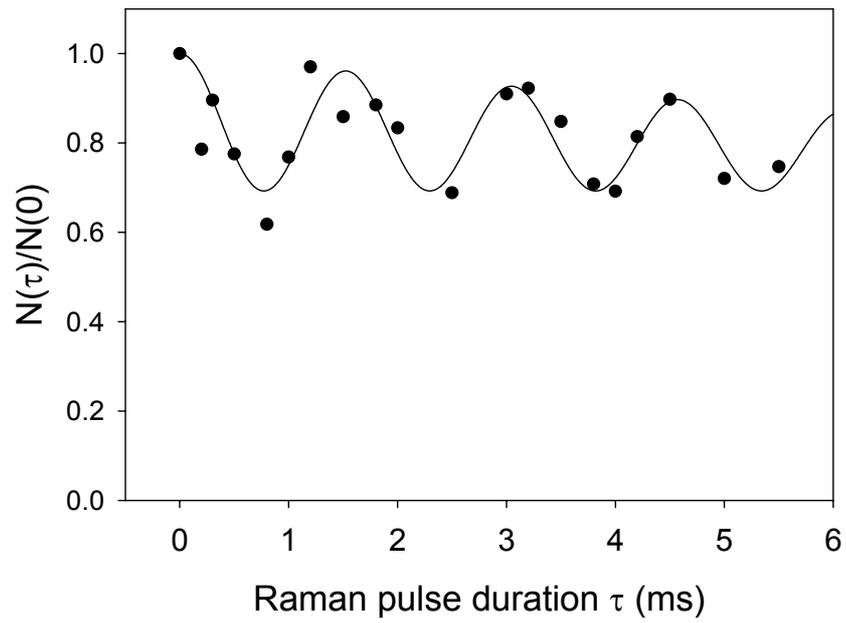

**Figure 4** Fraction of atoms remaining in the gas as a function of the duration of a Raman photoassociation pulse, showing oscillation between an atomic and a molecular gas.